\documentclass[nohyperref]{article}

\usepackage{microtype}
\usepackage{graphicx}
\usepackage{subfigure}
\usepackage{booktabs} 

\usepackage{hyperref}

\usepackage[accepted]{icml2022}

\usepackage{amsmath}
\usepackage{amssymb}
\usepackage{mathtools}
\usepackage{amsthm}

\usepackage[capitalize,noabbrev]{cleveref}

\theoremstyle{plain}

\theoremstyle{definition}

\theoremstyle{remark}

\usepackage[textsize=tiny]{todonotes}

\DeclareUnicodeCharacter{2212}{-}
\DeclareUnicodeCharacter{2009}{ }

\icmltitlerunning{Understanding chemical reactions via variational autoencoder and atomic representations}

\begin{document}

\twocolumn[
\icmltitle{Understanding chemical reactions via variational autoencoder and atomic representations}

\icmlsetsymbol{equal}{*}

\begin{icmlauthorlist}
\icmlauthor{Martin Šípka}{mff,prf}
\icmlauthor{Andreas Erlebach}{prf}
\icmlauthor{Lukáš Grajciar}{prf}
\end{icmlauthorlist}

\icmlaffiliation{mff}{Mathematical Institute, Faculty of Mathematics and Physics, Charles University, Sokolovská 83, 186 75 Prague, Czech Republic.}
\icmlaffiliation{prf}{Department of Physical and Macromolecular Chemistry, Faculty of Sciences, Charles University, Hlavova 8, 128 43 Prague 2, Czech Republic}

\icmlcorrespondingauthor{Martin Šípka}{martin.sipka@natur.cuni.cz}

\icmlcorrespondingauthor{Lukáš Grajciar}{lukas.grajciar@natur.cuni.cz}

\icmlkeywords{Collective variables, training on representation, chemical reactions}

\vskip 0.3in
]

\printAffiliationsAndNotice{} 


\begin{abstract}
On the time scales accessible to atomistic numerical modelling, chemical reactions are considered rare events. Atomistic simulations are typically biased along a low-dimensional representation of a chemical reaction in an atomic structure space, i.e., along the collective variable, to accelerate sampling of these improbable events. However, suitable collective variables are often complicated to guess due to the complexity of the transitions. Therefore, we present an automatic method of generating robust collective variables from atomic representation vectors, using either fixed Behler-Parrinello functions or representations extracted from pre-trained machine learning potentials. Variational autoencoder with these representations as inputs is trained while its latent space with arbitrary dimension gives us the set of collective variables. The resulting collective variables inherit all necessary invariances from the atomic representations and can be trained entirely unsupervised. The method's effectiveness is demonstrated using three different chemical reactions, one being the complex hydrolysis of a heterogeneous aluminosilicate catalyst. Lastly, we consider the method in the context of unseen atomic structure prediction, efficiently creating structures for different values of collective variables in a generative model fashion.
\end{abstract}

\section{Introduction}
\label{intro}
On the time scales accessible to atomistic numerical modelling, chemical reactions are considered rare events. Atomistic simulations are typically biased along a low-dimensional representation of a chemical reaction in an atomic structure space, i.e., along the collective variable, to accelerate sampling of these improbable events. In this paper, we propose a way of constructing a collective variable via machine learning methods. As a method of choice for dimensionality reduction, we use a variational autoencoder (VAE) \cite{vae}. Autoencoder has been used for the construction of collective variables in the past \cite{RAVE} but has been based on the use of handpicked descriptors such as distances, angles or even bare cartesian coordinates \cite{bglCVs}. We consider this approaches insufficient for the fully automatic collective variable identification. The handpicking of initial descriptors may be helpful when only one complex and/or well-known type of reaction is of concern but fails when we seek to fully automatize the process of reaction system exploration. Alternatively, the listing of all descriptors is doable for small systems, yet the total number of distances increases quadratically with the system size and a total number of angles cubically. Lastly, the cartesian coordinates do not respect the translational and rotational invariance and are not suitable except for some special cases. Our method brings novelty in the following:

\textbf{Supervised feature extraction:} We present a novel approach of training a variational autoencoder on top of a pre-trained model. This way, we simplify the challenging feature extraction task by a supervised pretraining of a machine learning potential. 
  
\textbf{Fully automatic:} Our method is truly fully automatic with few hyperparameters. It is, in nature, unsupervised, while supervised feature filtering might help when only limited data are available. 

\textbf{Linear scaling:} Representations are constructed using a finite cutoff, making the method suitable for large systems since the total size of all representation vectors scales linearly with the system size.

\textbf{Invariance:} We automatically take into account the rotational and translational invariance and periodicity since the representations respect them. 

\textbf{Computational time saving} Generation of our collective variables is integrated with the evaluation of the underlying machine learning potential. During simulation, we would calculate atomic representations only once and use them to get energy, forces and collective variables. Since the representation calculation is the most expensive part of the NNP evaluation, while the CV network is only a few layers deep, we get CVs basically for free.

\section{Building blocks of the representation-based variational autoencoder}
\subsection{Representations as inputs}
When working with molecular data, it is common to employ untrained representation functions or graph neural networks to reduce unstructured data to a fixed size vector. The vector is then called a representation vector and is a property of every atom in our system. A local environment of an atom and interactions with its neighbours are all captured to form input for predictions of all kinds (e.g. energy, charge, etc.). In our work, we employ two different approaches for generating representations. 

\textbf{ACFS}: For smaller systems we choose Behler-Parrinello Atom-centered symmetry functions (ACSF) \cite{BehlerACFS}, \cite{BPACFS}. While there are many modifications, we choose the original version of the atomic representation, which gives every atom type its own basis functions, in a one-hot encoding fashion based on an atom type. This way, ACSF can provide us with an untrained atomic representation of what environment every atom is in that can be used as input for our variational autoencoder in the next step.

\begin{figure}[ht]
\vskip 0.2in
\begin{center}
\centerline{\includegraphics[width=\columnwidth]{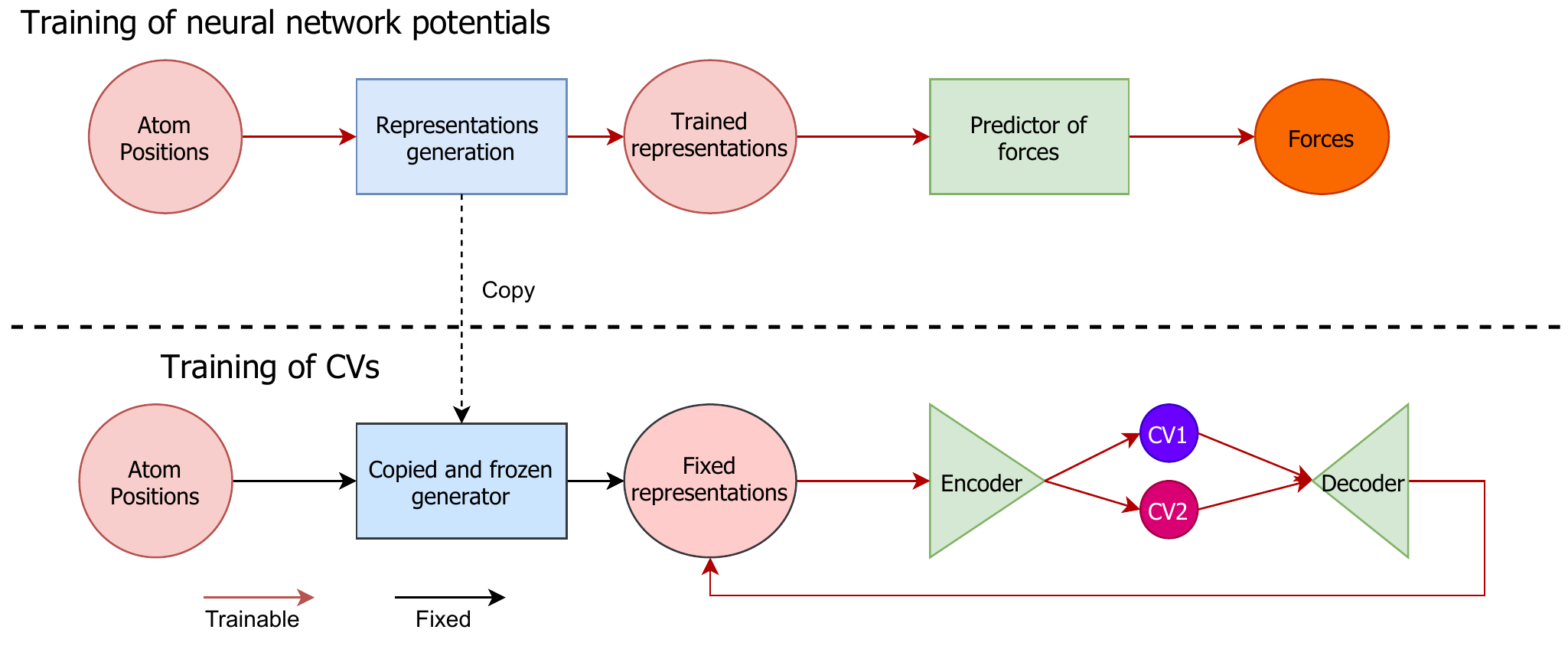}}
\caption{Training of our collective variables.}
\label{training}
\end{center}
\vskip -0.2in
\end{figure}

\textbf{Pretrained representations}: While the fixed representations are easy to use for every structure, we find that pretraining custom representations for our systems of interest can be beneficial for the collective variable training in the next stage. Pretraining is done by a supervised training of a neural network potential (NNP) on forces and energies obtained by accurate density functional (DFT) calculations. The process is as follows (\cref{training}):
\begin{enumerate}
  \item Collect a database of DFT calculations for the system of interest. Make the database as relevant as possible for the reaction we explore.
  \item Train a neural network potential on the database and test its performance. It is crucial that the potential is of good quality. We cannot fine-tune its layers during collective variable development and an inaccurate NNP will inevitably lead to very bad collective variables.
  \item Extract the atomic representation from the NNP, freeze all its weights and change all layers to evaluation regime (batch normalization, dropouts etc.)
\end{enumerate}
In our case, we choose to work with very important heterogeneous catalysts, zeolites, \cite{zeos} as our system of interest and a high-quality NNP was kindly provided to us by the authors of \cite{erlebachNNP}. The NNP is build using Schnet \cite{SCHNET} architecture. The reactions involving zeolites demonstrate that using tailored representations equipped with accurate DFT-level information makes it easier for autoencoder to understand the reaction and enables the construction of correct and relevant collective variables. 

\subsection{Representation arrangement and reduction}
There are numerous choices for arranging and ordering the final system representation vectors. One may take inspiration from the machine learning potentials, reducing first the representation and sum the reduced result. Alternatively, one may concatenate vectors atom by atom and construct one long vector, which is then the input of the autoencoder. Many more complicated variants exist and are worth exploring in the future. However, we choose the simple vector concatenation approach as it works best for our purposes. 

Our simple concatenating approach comes with a disadvantage. The resulting representations are rather long. For our typical representation model of aluminosilicates that outputs 128 components and a system of 39 atoms, we get, in total, almost 5000 components (features). This may be unnecessary. Therefore, we prefilter our inputs for larger systems and bigger models, taking only features that are most different between reactants and products. A simple difference of a feature means for reactants and products serves as a good measure of feature importance. We must note this is a supervised element in an otherwise unsupervised method, but since we have a separate dataset for reactants and products, we can use it as it is easy to label the data. For smaller systems, as our first example reaction of keto-enol tautomerism, this step is not required and is omitted.

Finally, we note that the choice of different encoder architecture strongly influences the behaviour and invariance of a collective variable with respect to the swapping of atoms of the same kind. We plan to investigate the influence of encoder architecture in the future.

\subsection{Variational autoencoder}
Our variational autoencoder trains with representations as inputs. This means it reduces the pre-computed representations to form a latent space and then tries to reconstruct the representation vectors as precisely as possible.

For an encoder, just one linear layer without activation is viable for simpler reactions. We can use two hidden layers with shifted softplus activation for more complex reactions. The shifted softplus activation function was chosen since it is successfully used in Schnet architecture \cite{SCHNET}. The latent space of the variational autoencoder contains as many neurons as many collective variables we wish to use. The minimalistic encoder was used mainly because we want to avoid its overfitting as much as possible. We need to keep in mind that since we are interested in the CV values as well as in its derivatives, overfit and highly oscillating CVs would give us very steep gradients for some values of the representation vectors. Avoiding overfitting is also a reason for using $L2$ regularization. An $L2$ value of $10^{-6}$ or $10^{-4}$ worked well in our experiments. 

Our decoder can be a deeper network than the encoder since its derivatives are not used in simulations. Actually, the decoder itself is only used for training purposes and is not used in simulation runs at all. Therefore, we use between five and seven hidden layers, starting with 16 hidden neurons, doubling until we reach the size of the representation that we are trying to match during training. ReLU activation function was used in this case. 

Full training parameters for all reactions are reported in the Appendix \ref{hparams}.

\section{Training}
We, based on whether we have a good pre-trained representation, follow two possible paths to train the collective variables. In the following paragraph, we refer to the datapoint as one simulations step, in other words, a single geometry that is saved during our calculations (see Appendix \ref{compdat} for more details on the computational set-up used to generate structural datapoints). The first step is always to collect "good" equilibrium data for our reaction's initial and final state. By "good", we mean that rather than just simulating as many data points as possible, we focus on having all relevant events and atom motions for equilibrium products and reactants captured. We simulate at different temperatures to cover the wider distribution of energies and choose the structure saving frequency large enough such that the subsequent data points do not correlate too much. With this database ready, we move to a second step. 
\begin{itemize}
  \item If no pre-trained representation is available, we calculate the ACFS vectors for every atom in every data point in our database. We use periodic boundary conditions at this step if our simulation is periodic and end up with a fixed-size vector that we save for further training. 
  \item In case we have a good machine learning potential available, we extract the representation network from the potential and calculate the representations for every atom and every datapoint using this model. Periodic conditions are automatically taken care of by the model, and the resulting fixed-size vector is saved. 
\end{itemize}
We precalculate and save our representation since we do not modify it afterwards, and this calculation is the most time-consuming training step. 

In the next step, we define our variational autoencoder. We choose the number of collective variables we would like to use and set the hyperparameters of the training. After training, we take the value of our CV for every point in our dataset and rescale the collective variable, so that is it in $[-1,1]$ interval. This way, we can, e.g.,  store the bias on a grid, making the metadynamics \cite{mtd} simulations more efficient \cite{plumed}. After saving the frozen trained encoder model, we can proceed with a biased simulation.

\section{Experiments}
We demonstrate the collective variables on three reactions. We choose the keto-enol tautomerization for the first one, a simple reaction of a simple molecule in a gas phase. We use plain ACFS representations and show that we can train a robust collective variable for such a system, allowing us to bias the reaction and obtain a free energy surface. Next, we simulate a hydrogen jump in a much more complex environment in an aluminosilicate porous material called chabazite. In this case, we show that a pre-trained high-quality representation allows us to get the desired CV and free energy profiles of a reaction, while with ACFS, we struggle to do so. The last system we investigate is the hydrolysis of a chabazite zeolite. A free molecule of water is present in the chabazite cell, and it reacts with the aluminosilicate. We use the same pre-trained Schnet-like representations as in the previous example and obtain a collective variable to simulate the reaction. 
\subsection{Keto-enol tautomerization} \label{ketochapter}
For the first reaction, we choose the keto-enol tautomerization. Using  ACFS representations, we show that we can train a robust collective variable for such a system, allowing us to bias the reaction and obtain an estimate of a free energy surface (FES).

\begin{figure}[ht]
\vskip 0.2in
\begin{center}
\centerline{\includegraphics[width=\columnwidth]{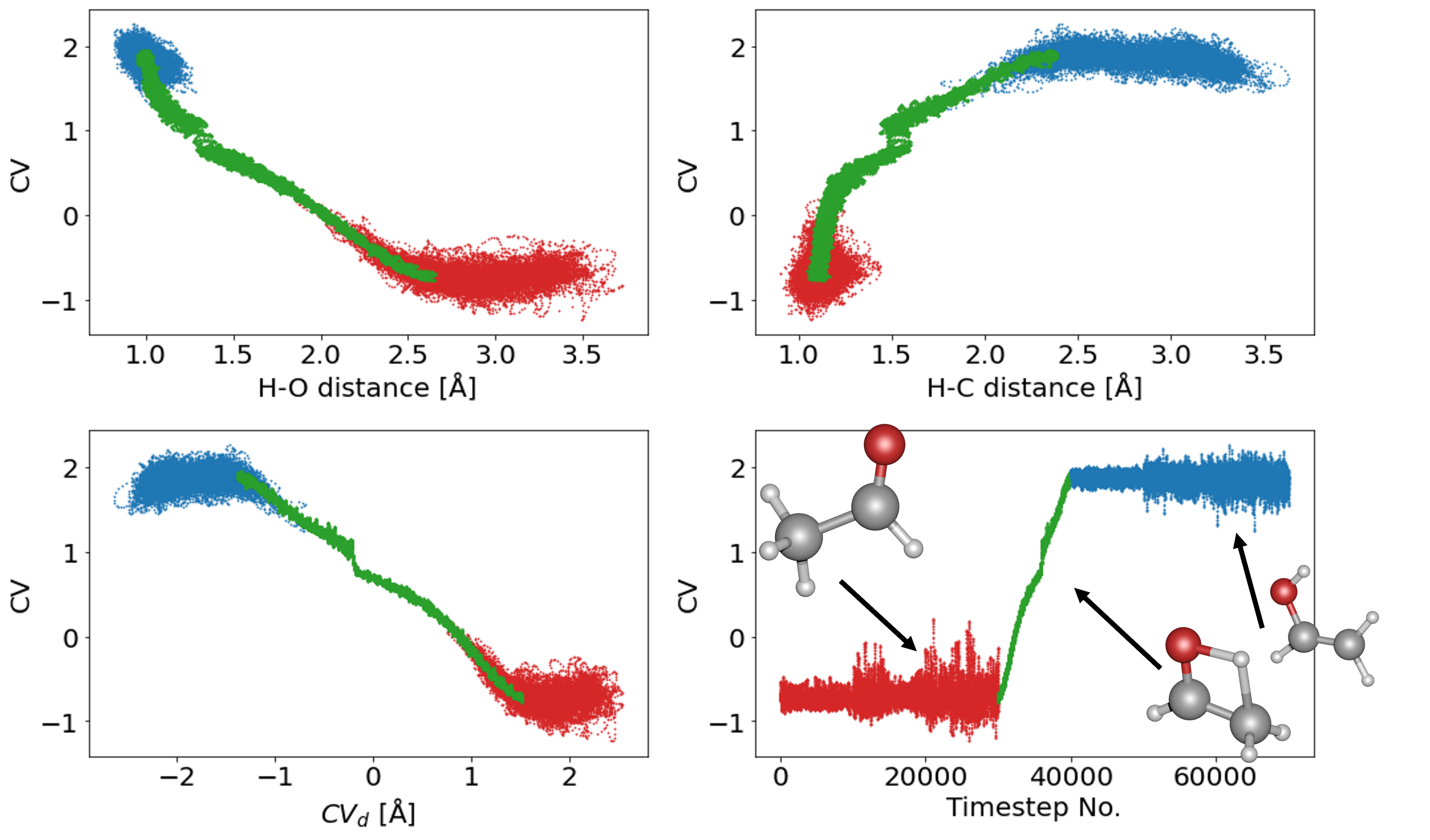}}
\caption{Projections of our collective variables to important distances. Red points represent reactants, blue points represent products and green ones are slow-growth benchmark dynamics using $CV_{d}$ \eqref{ketoenoldistcv}. In the bottom-right Figure, we concatenate all datasets, putting reactants first slow-growth in the middle and products last, while keeping the order of simulation steps. We then plot the evolution of a collective variable on this simulated datapoints. Training was done with only blue and red points.}
\label{keto_enol_project}
\end{center}
\vskip -0.2in
\end{figure}

In the reaction hydrogen H jumps from carbon C to oxygen O, changing an acetaldehyde to a vinylalcohol (see \cref{keto_enol_project}). Denoting a H-O distance by $d_{H-O}$ and H-C distance by $d_{H-C}$, we can construct a good collective variable in the form
\begin{equation}
    CV_{d} = d_{H-C} - d_{H-O}.
\label{ketoenoldistcv}
\end{equation}
We will use this collective variable as a reference and also generate a slow-growth (SG) dynamics \cite{slowgrowth} using this $CV_{d}$, which connects reactants to products via a transition state. These SG dynamics were not used during training, but they serve as a good benchmark for our collective variable even before running the simulation. We would expect that our collective variable would change close to linearly with $CV_{d}$ along SG dynamics. We can indeed see this is the case in the \cref{keto_enol_project}. 

\begin{figure}[ht]
\vskip 0.2in
\begin{center}
\centerline{\includegraphics[width=\columnwidth]{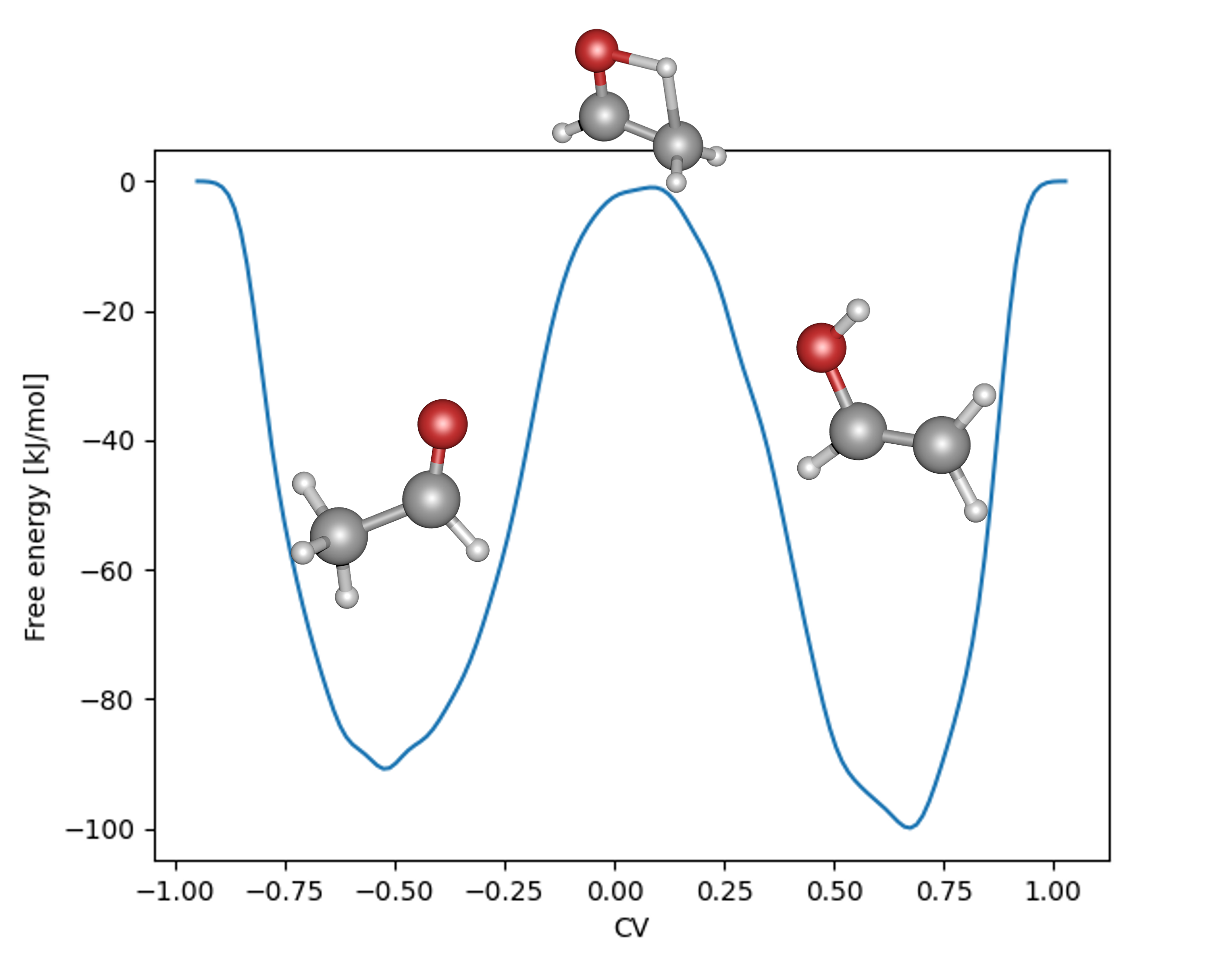}}
\caption{Rough free energy surface for keto-enol transition.}
\label{fesketoenol}
\end{center}
\vskip -0.2in
\end{figure}

Using standard metadynamics \cite{mtd} with the gaussian height set to $1.2 KJ/mol$ and its width to $0.04$ we also calculate the free energy surface of the reactions. We use a relatively coarse setting since we do not aim for the precise barrier estimation. Finer FES could be obtained by: i) fine-tuning the standard metadynamics parameters, or ii) using more advanced methods such as well-tempered metadynamics \cite{wtmtd}, or the OPES method \cite{opes}. However, we instead use metadynamics to showcase that our method produces robust collective variables, readily deployable for practical applications. FES can be seen in the \cref{fesketoenol}, and it qualitatively agrees with previous static DFT-based calculations for keto-enol tautomerization of 2- and 4-hydroxyacetophenone \cite{ketoenol}. For the metadynamics, we used modified PLUMED software \cite{plumed}, \cite{plumed2}, while for the evaluation of the forces, we employed the PM6 model from the package CP2K \cite{cp2k}.

\subsection{Hydrogen jump in chabazite} \label{hjumpchapter}
\begin{figure}[ht]
\vskip 0.2in
\begin{center}
\centerline{\includegraphics[width=\columnwidth]{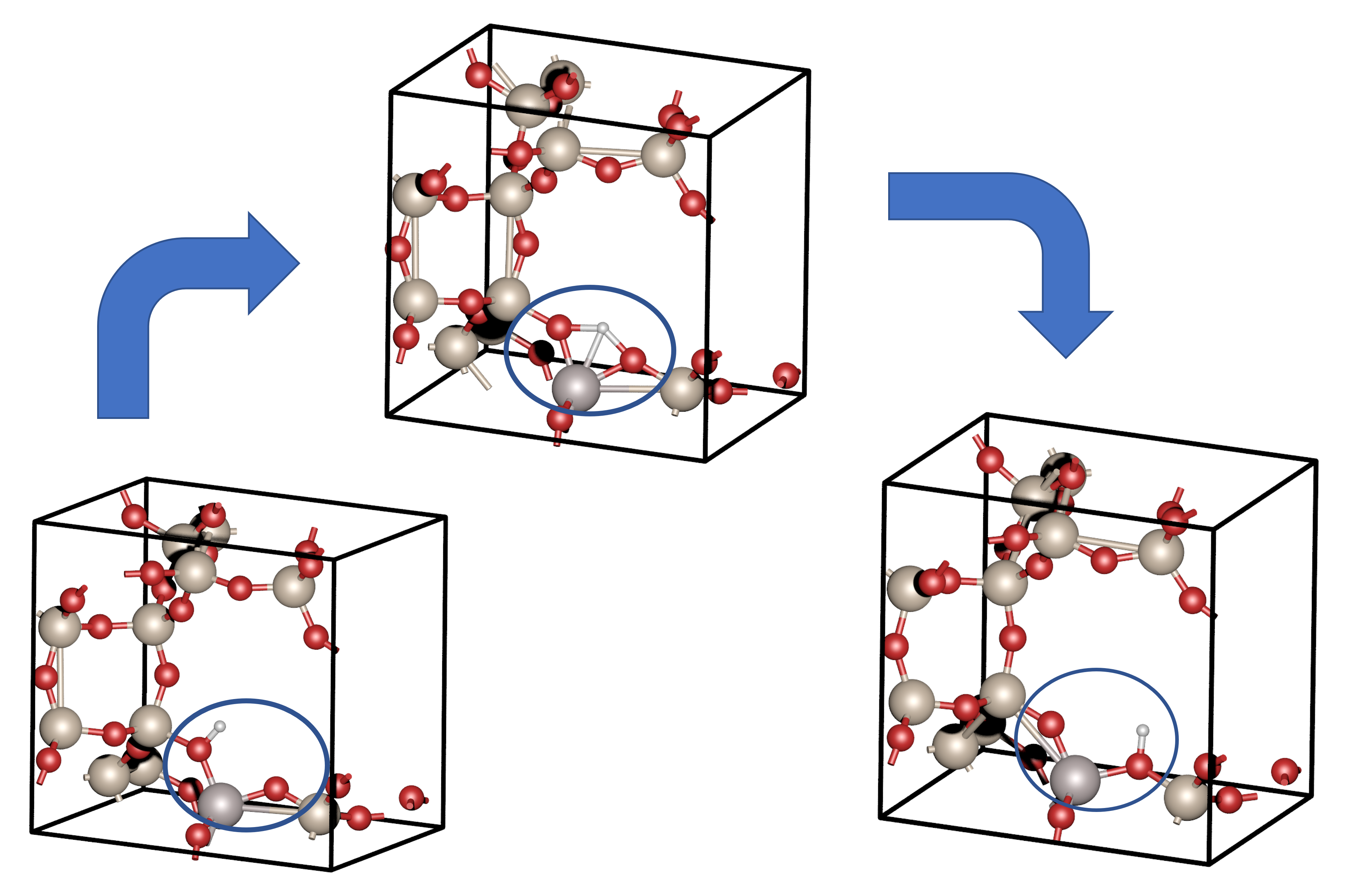}}
\caption{Illustration of the hydrogen jump in a chabazite unit cell (with a single aluminium in the unit cell).}
\label{hjumpillustrate}
\end{center}
\vskip -0.2in
\end{figure}

The next reaction we investigate is also a hydrogen jump (\cref{hjumpillustrate}), except this time in a much more complex system, in an aluminosilicate zeolite chabazite (CHA) \cite{zeodat}, using a neural network potential (NNP) for force and energy evaluation and the NNP's atomic representations to generate collective variables. This is done using libraries Schnetpack \cite{schnetpack}, and ASE \cite{asecitation} with minor modifications allowing us to use a collective variable defined in the ASE environment in PLUMED.

\begin{figure}[ht]
\vskip 0.2in
\begin{center}
\centerline{\includegraphics[width=\columnwidth]{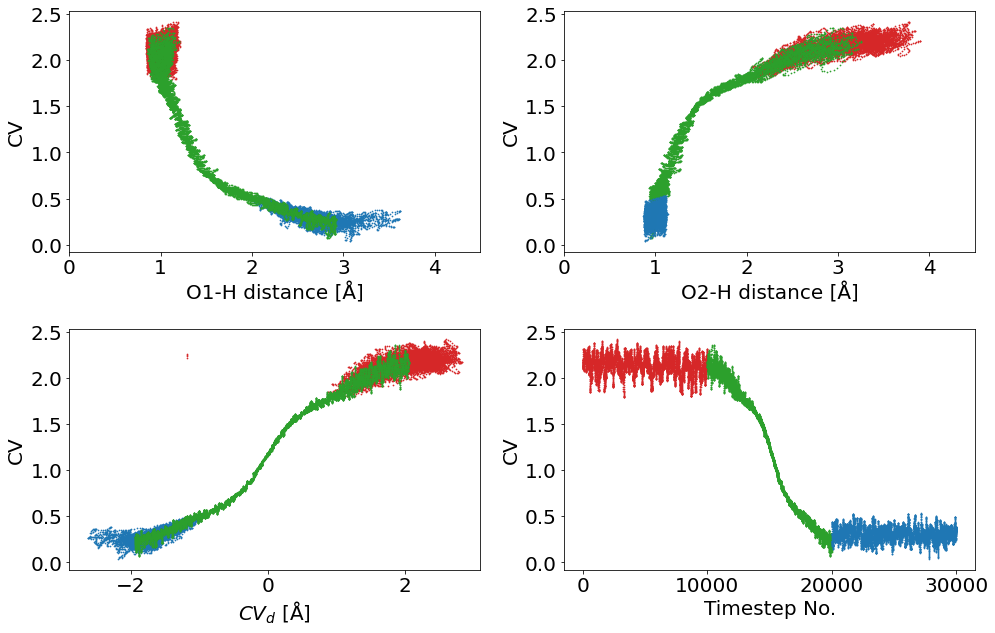}}
\caption{Projections of our collective variables to important interatomic distances. Red points represent reactants, blue points represent products and green ones are slow-growth dynamics using $CV_{d}$ \eqref{hjumpdistcv}. Training was done with only blue and red points. The bottom-right figure is yet again the evolution of a collective variable on a concatenated dataset of equilibrium and slow-growth simulations}
\label{hjumpprojections}
\end{center}
\vskip -0.2in
\end{figure}

\begin{figure}[ht]
\vskip 0.2in
\begin{center}
\centerline{\includegraphics[width=\columnwidth]{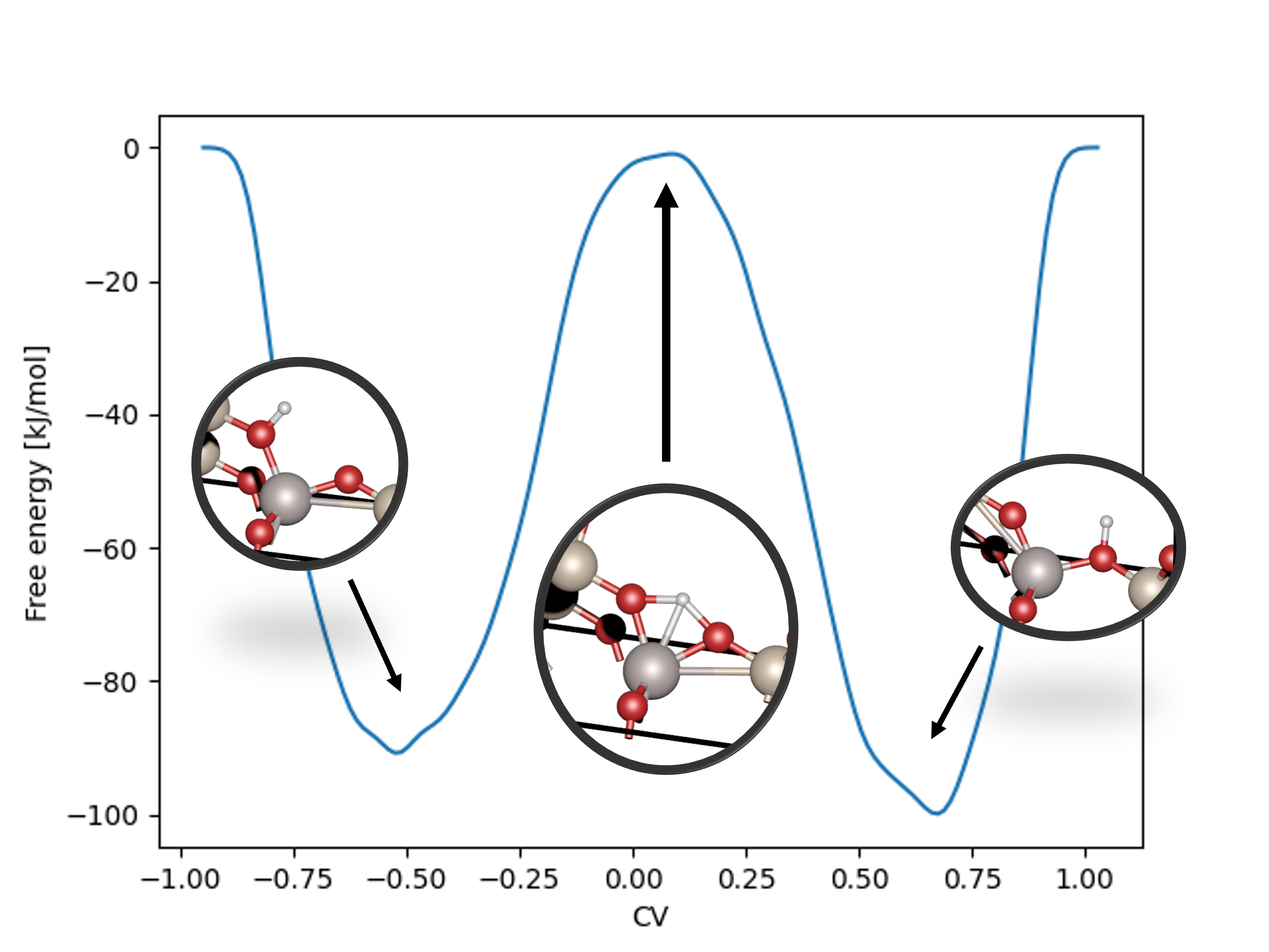}}
\caption{Coarse free energy surface for the chabazite hydrogen jump.}
\label{hjumpfes}
\end{center}
\vskip -0.2in
\end{figure}

Similarly as in the previous reaction, we identify the important interatomic distances to focus on. The hydrogen jumps from one of the oxygens labeled $\text{O}_1$, to another oxygen labeled  $\text{O}_2$. The standard distance-based CV can then be constructed as
\begin{equation}
CV_{d} = d_{H-O_1} - d_{H-O_2}.
\label{hjumpdistcv}
\end{equation}
We can generate the slow-growth trajectory using the $CV_{d}$ and yet again look at the projections (\cref{hjumpprojections}). We reiterate that we do not use the slow-growth trajectory for the training of our CVs. Using standard metadynamics with the same parameters as in the previous reaction, we can also estimate the FES of the reaction \cref{hjumpfes}. The reaction barriers from the estimated FES are well within the previously reported DFT-based proton jump barriers in aluminosilicate CHA, which were shown to span the energy range of approx. 70-110 kJ/mol \cite{hjump}. 

\subsection{Chabazite hydrolysis} \label{hydrochapter}
\begin{figure}[ht]
\vskip 0.2in
\begin{center}
\centerline{\includegraphics[width=\columnwidth]{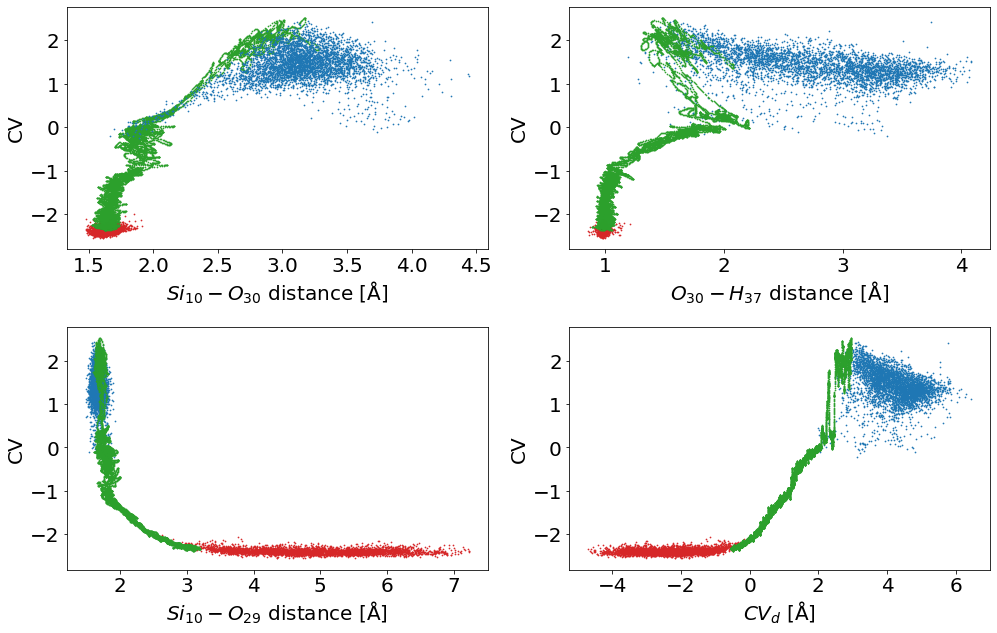}}
\caption{Chabazite hydrolysis and its projection to a collective variable described in \ref{hydrocv}}
\label{hyroprojections}
\end{center}
\vskip -0.2in
\end{figure}

The last reaction of interest is a purely siliceous chabazite hydrolysis \cite{hydrocha}. A water molecule freely moving in chabazite channels binds to a silicon atom, splits, breaks one of the frameworks Si-O bonds and generates two SiOH (silanol) groups. We employ the same computational set-up and NNP as for the hydrogen jump (see \ref{hjumpchapter}). We yet again define a benchmark CV (see \ref{hydrocvsec} for the definition). The projections to relevant distances are depicted in the \cref{hyroprojections}. 

For this reaction, we do not report a free energy surface. Instead, we use a method called steered dynamics \cite{steeredD} and push the reaction in the desired direction. The method works by adding a moving bias potential in the form
\begin{equation}
U(CV, t) = \frac{1}{2}\kappa(CV - C(t))^2
\end{equation}
where $\kappa$ is, in our case, a fixed constant, the steepness of the bias potential to the simulation, while $C(t)$ is the time-dependent desired CV value. 

For our hydrolysis case, we set a relatively high $\kappa = 30 \ 000$ and in approximately $6 \  \text{ps}$ with a stepsize of $0.5 \  \text{fs}$ we move from reactants to products. In the cref{hydrocvsec} we show the evolution of our collective variable and a structure corresponding to the CV value of 0. This structure is expected to be close to the transition state. 
\begin{figure}[ht]
\vskip 0.2in
\begin{center}
\centerline{\includegraphics[width=\columnwidth]{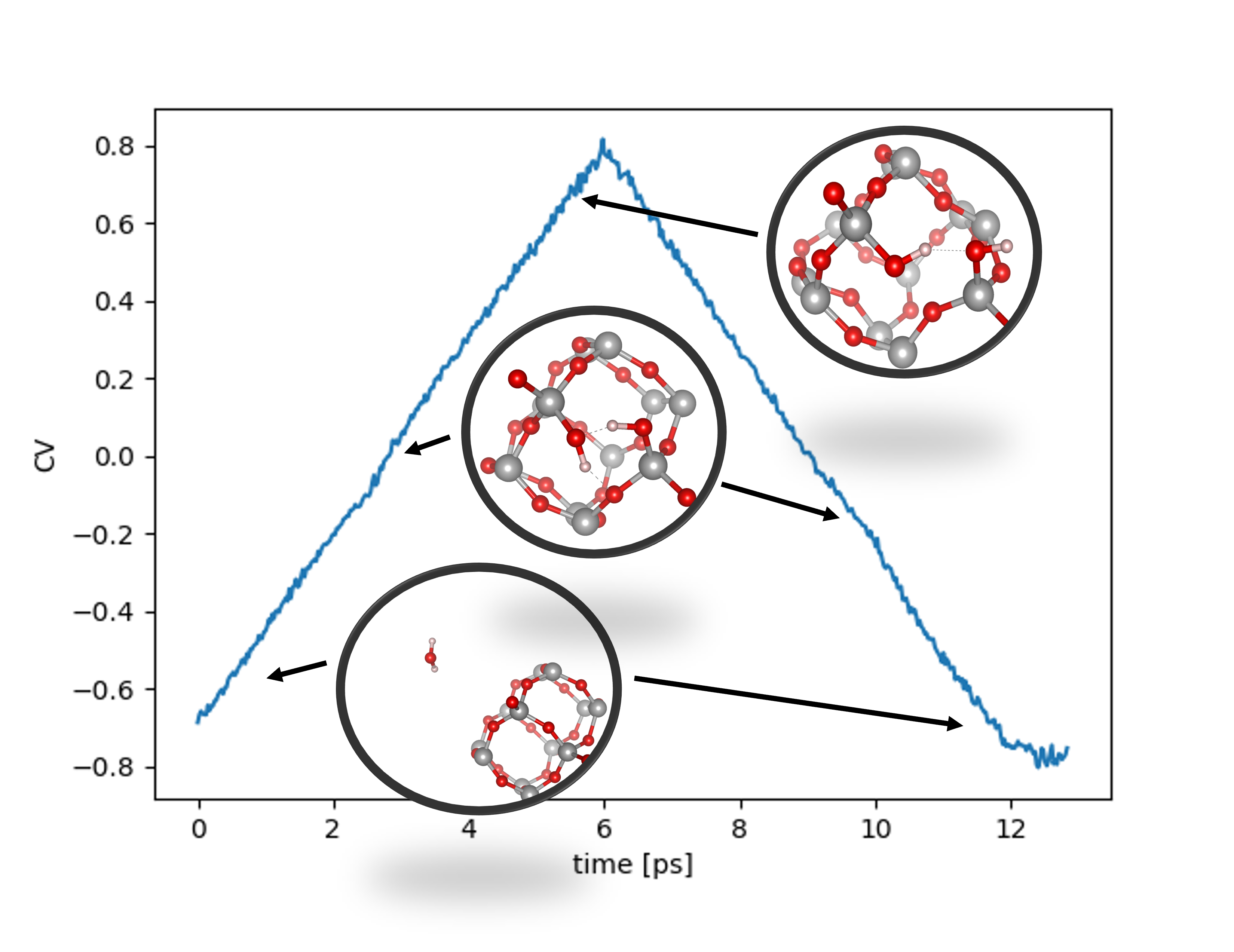}}
\caption{Chabazite hydrolysis: evolution of the trained collective variable using moving restrain technique. For low CV values we have reactants, for high values we get products.}
\label{hydrocv}
\end{center}
\vskip -0.2in
\end{figure}

\textbf{Generative modelling perspective:} We will now consider the hydrolysis reaction from a different point of view. There have been numerous works in generative neural network models in chemistry (see, e.g. comprehensive review \cite{generativeReview}). The idea is that we predict a structure for a given property value. A specialized decoder network has achieved this in the reviewed papers. However, decoding a fixed size vector into a realistic structure is challenging. 

\begin{figure}[ht]
\vskip 0.2in
\begin{center}
\centerline{\includegraphics[width=\columnwidth]{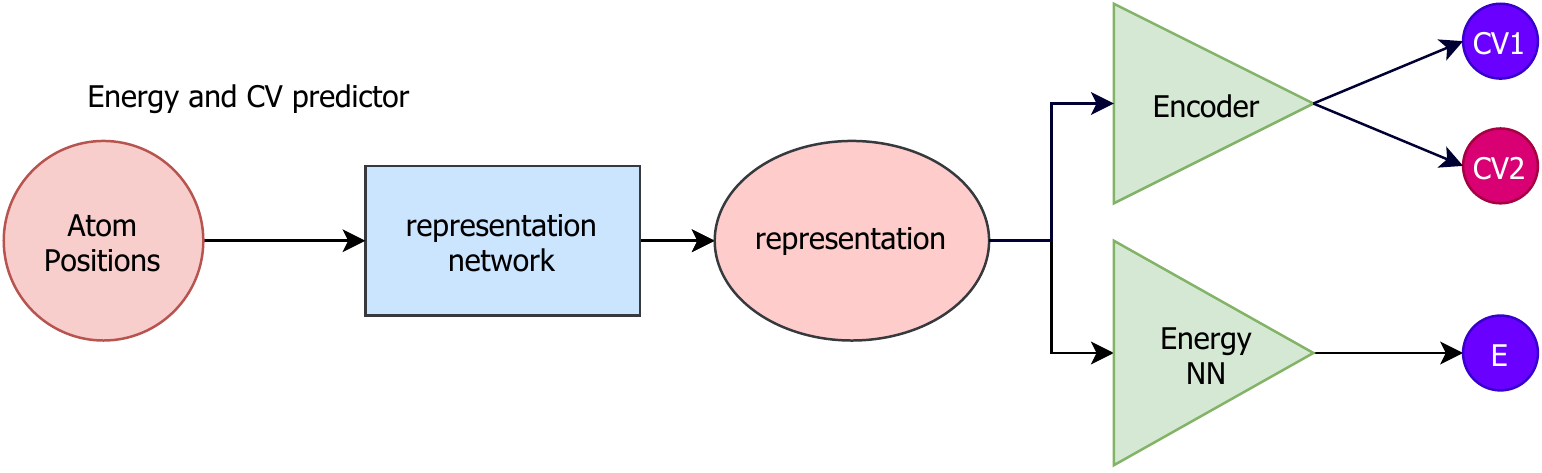}}
\caption{Single neural network for energy and collective variable prediction.}
\label{generator}
\end{center}
\vskip -0.2in
\end{figure}

We demonstrated that a structure close to our transition state could be generated differently in our last reaction. By starting with some known reactant structure, we iterate and move the value of a reaction coordinate to where we believe the transition state is. An advantage, in this case, is that the neural network potential is keeping our steps guided within realistic and permissible energy values. Our two neural networks together (\cref{generator}) form an iterative mechanism capable of generating unseen structures as a decoder would. So far, it has been tested just on a single reaction, but increasing latent space and training on more known structures are planned.
\section{Conclusion}
In this paper, we introduced a novel method of automatic collective variable identification employing variational autoencoder and pre-trained neural network potential representations. It is characterized by supervised feature extraction that circumvents the hard task of handpicking good structure descriptors. It is linear scaling with the system's size, invariant to structure rotations and translations, and efficiently merges with the newly developed neural network potentials. We demonstrated its capabilities in three different reactions, from simple keto-enol tautomerization to complex heterogeneous catalyst hydrolysis. Our collective variables agreed well with expertly identified reaction coordinates for every example. The method proved itself on small molecules using fixed, untrained representations and on larger periodic structures using Schnet pre-trained representations. Lastly, we discussed the method's potential in the context of generative structure modelling and outlined our development plans. 
\bibliography{example_paper}
\bibliographystyle{icml2022}

\newpage
\appendix
\onecolumn
\section{VAE Training hyperparameters}\label{hparams}

\begin{tabular}{ |p{1.5cm}||p{1.4cm}|p{1.4cm}|p{1.4cm}|p{1.5cm}|p{1.5cm}|p{1.5cm}|p{1.5cm}|p{1.5cm}|  }
 \hline
 \multicolumn{9}{|c|}{Hyperparameters} \\
 \hline
 Reaction& Rep. used& Rep. size& No. atoms& features selected&  epochs& Init. LR& L2 regularization& Batch size \\
 \hline
 keto-enol   & ACFS   & 36 & 7 & All (252) & 15 & 0.001 & $10^{-6}$ & 30\\
 H-jump      & Schnet & 128 & 37 & 100 / 4736 & 20 & 0.0005 & $10^{-4}$ & 20 \\
 Hydrolysis  & Schnet & 128 & 39 & 300 / 4992 & 15 & 0.0005 & $10^{-4}$ & 20 \\
 \hline 
\end{tabular}

In all cases we used StepLR learning rate scheduler, implemented in PyTorch, with $\text{step\_size}=1$ and $\gamma=0.95$. For keto-enol tautomerism reaction, we used 12 distance ACFS and we did not use angle ACFS.

\subsection{Variational autoencoder architecture}
\begin{tabular}{ |p{1.5cm}||p{1.5cm}|p{1.5cm}|p{2cm}|p{1.5cm}|p{1.5cm}|p{3.2cm}|}
 \hline
 Reaction& Inputs& Encoder layers& Encoder neurons & latent dimension& Decoder layers& Decoder neurons \\
 \hline
 keto-enol   & 252   & 1 & 1 & 1 & 3& 16-32-252 \\
 H-jump      & 100 & 3 & 100-50-1 & 1 & 5 & 16-32-64-128-100 \\
 Hydrolysis  & 300 & 3 & 300-150-1 & 1 & 6 & 16-32-64-128-256-300 \\
 \hline 
\end{tabular}
\subsection{Data}
Here, we report the amount of data we used for the training for every reaction. 
\textbf{Keto-enol tautomerism}
We use 10 000 timesteps with a temperature of 300K, 10 000 timesteps of 1000K and 10 000 timesteps of 2000K equilibrium data for reactants and products. 
\textbf{Hydrogen jump}
We use 10 000 timesteps of 300K equilibrium data for reactants and products. 
\textbf{hydrolysis}
For this reactio we try to get a rich dataset using for both reactans and products:
\begin{itemize}
  \item 1000 timesteps of 300Kequilibrium data
  \item 1500 timesteps of 1000K equilibrium data
  \item 1500 timesteps of data, gradually heated from 450K to 2000K
\end{itemize}
In total 4000 timesteps for reactants and 4000 timesteps for products. 

\section{Benchmark CV for hydrolysis reaction}\label{hydrocvsec}
\begin{figure}[ht]
\vskip 0.2in
\begin{center}
\centerline{\includegraphics[width=\textwidth*2/3]{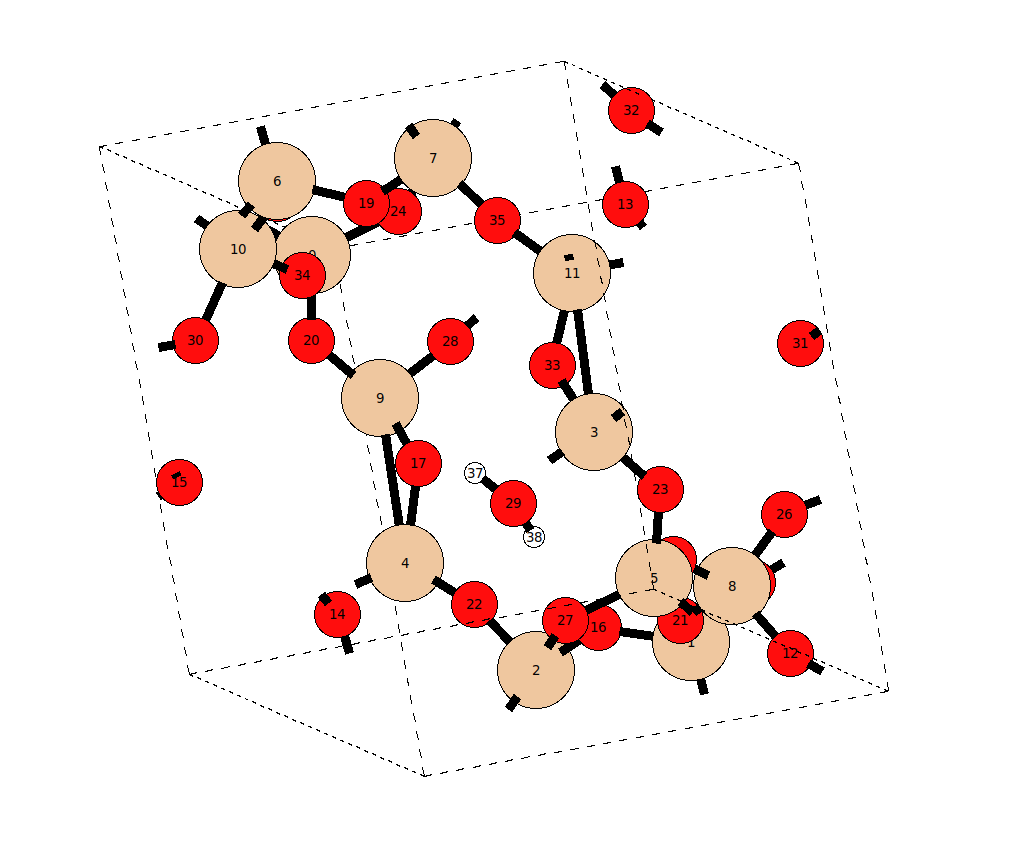}}
\caption{Reactant in our hydrolysis reaction. The atomic indices are drawn to better understand the collective variable used.}
\label{hydroreact}
\end{center}
\vskip -0.2in
\end{figure}

Let us consider the hydrolysis of an aluminosilicate from \ref{hydrochapter}
In reactants (see the \cref{hydroreact}) we have a a freely moving water with oxygen $O_{29}$, and hydrogens $H_{37}$ and $H_{38}$. There are two  Silica atoms, $Si_{10}$ and $Si_{11}$ connected by an oxygen $O_{30}$. Out goal is to break the bonds $Si_{10} - O_{30}$ and $O_{29} - H_{38}$ and form bonds $Si_{10} - O_{29} - H_{38}$ and $Si_{11} - O_{30} - H_{37}$ (see \cref{hydroreact}). Therefore we can define a $CV_d$ as
\begin{equation}
    CV_d = d_{Si_{10}-O_{30}} + d_{O_{29}-H_{37}} - d_{Si_{10}-O_{29}}
\end{equation}
\section{Reference structure models and methods}\label{compdat}
The model for the zeolite is a 12 T site periodic cell of chabazite (CHA) zeolite \cite{zeodat}, which contains one microporous cage ($\text{a} = \text{b} = \text{c} = 9.27$ Å, $\alpha = \beta = \gamma =$ 96°). CHA has one T site to which four inequivalent oxygen atoms are connected. To generate structural data for keto-enol tautomerization we employed a cubic unit cell with the side length of 15 Å. 

The reactant and product ab initio molecular dynamics and dynamical slow-growth calculations for all reactions in the article, carried out in order to obtain reference structures, were performed using density functional theory (DFT) as implemented in the VASP 5.4. code \cite{vasp}, with the exchange-correlation functional of Perdew–Burke–Ernzerhof \cite{pbe}, and a D3 dispersion correction \cite{dftd3}. Wavefunctions are described by a plane-wave basis with a kinetic energy cutoff of 400 eV. The electronic structure was calculated at the gamma point only, which is sufficient if only structure generation is of importance. The ab initio molecular dynamics simulations were performed in the NVT ensemble, with temperature controlled by the Nosé–Hoover thermostat.  
\end{document}